\documentclass[aps,prb,amsfonts,amssymb,twocolumn,amsmath,draft,floatfix,showpacs]{revtex4-1}
\usepackage{amsbsy}
\usepackage[dvips]{color}
\usepackage[dvips]{graphics}
\usepackage{graphicx}% Include figure files
\usepackage{bm}% bold math

\newcommand{\be}{\begin{equation}}
\newcommand{\ee}{\end{equation}}
\newcommand{\bea}{\begin{eqnarray}}
\newcommand{\eea}{\end{eqnarray}}

\DeclareMathOperator{\sech}{sech}
\DeclareMathOperator{\csch}{csch}

\allowdisplaybreaks

\begin{document}

\title{Probing interfacial pair breaking in tunnel junctions based on \\
the first and the second harmonics of the Josephson current}

\author{Yu.\,S.~Barash}
\affiliation{Institute of Solid State Physics, Russian Academy of Sciences,
Chernogolovka, Moscow District, 142432 Russia}

%\date{\today}

\begin{abstract}
It will be shown that a pronounced interfacial pair breaking can be
identified in Josephson tunnel junctions provided the first $j_{1c}$
and the second $j_{2c}$ harmonics of the supercurrent, as well as
the depairing current in the bulk $j_{dp}$, are known. Namely,
within the Ginzburg-Landau theory a strong interfacial pair breaking
results in the relation $j_{2c}j_{dp}\gg j_{1c}^2$, while in
standard junctions, with negligibly small pair breaking, the
relation of opposite character takes place.
\end{abstract}


\maketitle

\section{Introduction}

A remarkable property of superconducting weak links is that the local
conditions in a small transition region control the whole process of
charge transport. For the same reason, interface-induced
suppression of the superconducting condensate density can have a
considerable influence on the Josephson effect. A strong surface pair
breaking has been theoretically
established for various unconventional superconductors as well
as for magnetic interlayers under certain conditions
\cite{Buchholtz1981,Barash1995,Shiba1995a,*Shiba1995b,*Shiba1995c,%
*Shiba1996,Nagai1995,Sauls1995a,*Sauls1995b,Alber1996,Agterberg1997,%
Sauls1988,Nazarov2009}. Therefore, probing the condensate
density near the interface would provide valuable information for
studying and controlling fundamental characteristics of the
superconducting junctions. It is still an ongoing problem for the
junctions though the order parameter profiles near superconductor-vacuum
surfaces have been recently determined using a Scanning Tunneling Microscopy
method with a superconducting tip \cite{Dynes2009}.

In superconducting tunnel junctions the first harmonic $j_1=j_{1c}
\sin\chi$ usually strongly dominates the Josephson current $j=j_{1c}
\sin\chi+j_{2c}\sin2\chi+\dots$, while the second harmonic $j_2=
j_{2c}\sin2\chi$ represents a small correction to the first one,
$|j_{2c}|\ll|j_{1c}|$, mostly due to a small junction transparency.
Qualitatively different phase dependencies of the two harmonics
allow to study and distinguish between them experimentally. It is
therefore of interest to find out, which characteristic properties of
the superconducting junctions can be identified with the data provided
by the two harmonics. Thus the first harmonic, as opposed to the second
one, is known to be noticeably suppressed both at $0$-$\pi$ transitions
as well as in the junctions involving unconventional superconductors
with special interface-to-crystal orientations \cite{Il'ichev1999,%
Lindstrom2003,Lindstrom2006,Mannhart2002,Golubov2004,Ryazanov2004,%
BuzdinPRB2005,Goldobin2007}.
However, except for these special cases, the relation $|j_{2c}|\ll|j_{1c}|$
always takes place and does not qualitatively discriminate between
various superconducting tunnel junctions.

This paper suggests a test, which will be derived within the
Ginzburg-Landau (GL) theory and will allow identification of a
pronounced interfacial pair breaking in tunnel junctions, provided
the first and the second harmonics, as well as the depairing current
in the bulk $j_{\text{dp}}$, are known.
The relation $j_{2c}j_{\text{dp}}
\gg j_{1c}^2$ will be shown to take place in tunnel junctions with a
strong interfacial pair breaking, while in standard
tunnel junctions, with negligibly small pair breaking, it will be
$j_{2c}j_{\text{dp}}=0.27j_{1c}^2 < j_{1c}^2$. The specific
temperature dependencies of the two harmonics near $T_c$ will also
be determined. The self-consistency is shown to alter
existing estimates of both harmonics considerably.
Initially, the theory is based on the interface free energy,
containing only the terms, which are quadratic or bilinear in the
superconducting order parameters. Later on, the study will be extended
to include the next order, quartic and biquadratic,
corrections. They will be shown to result in the material dependent
coefficients, which are independent of temperature near $T_c$ and
should, generally, be kept on a par with numerical terms of the order
of unity in the expressions for the order parameters.

\section{Basic equations}

Consider tunnel junctions with the spatially constant width, which is
much less than the Josephson penetration length, and with a plane
interlayer at $x=0$ of zero length within the GL approach. Assume the
usual form of the GL free energy, which
applies, for example, to $s$-wave and $d_{x^2-y^2}$-wave junctions:\,
${\cal F}={\cal F}_{b1}+{\cal F}_{b2}+{\cal F}_{int}$. Here ${\cal
F}_{b1(2)}$ are the bulk free energies of two superconducting leads
and ${\cal F}_{int}$ is the interface free energy. For a junction
with two identical superconductors, the bulk free energies have
identical coefficients
\begin{multline}
{\cal F}_{b1(2)}=\!\!\int\nolimits_{V_{1(2)}}\!\Bigl(
K\left|\pmb\nabla\Psi_{1(2)}\right|^2+
a\left|\Psi_{1(2)}\right|^2+\\ +
(b/2)\left|\Psi_{1(2)}\right|^4\Bigr)dV_{1(2)}.
\label{fb1}
\end{multline}
Here $K,b,\alpha>0$, $a=\alpha\tau=-\alpha (T_c-T)/T_c$.

Asymmetry can be generally maintained by different conditions on
the opposite sides of the interface, as in $d$-wave junctions with
different crystal-to-interface orientations, and/or in junctions with
asymmetric magnetic interfaces. Then the interface free energy
incorporates different contributions from the two superconducting banks:
\begin{eqnarray}
{\cal F}_{\text{int}}=\!\int\nolimits_{S}\Bigl[g_{11}
\left|\Psi_{1}\right|^2+
(1/2)h_{11}\left|\Psi_1\right|^4+g_{22}\left|\Psi_2\right|^2+ \nonumber \\
(1/2)h_{22}\left|\Psi_2\right|^4+
h_{12}\left|\Psi_1\right|^2\left|\Psi_2\right|^2
+\bigl(g_{12}+\eta_{1}\left|\Psi_1\right|^2+ \nonumber \\
+\eta_{2}\left|\Psi_2\right|^2\bigr)\left|\Psi_1-\Psi_2\right|^2 +
f_{12}\left|\Psi_1^2-\Psi^{2}_2\right|^2\Bigr]dS\,.\,\,
\label{fint1}
\end{eqnarray}

In addition to the main terms, which are quadratic or bilinear in the order-parameter
moduli, the quartic and biquadratic terms of the next order of smallness near $T_c$
are kept in \eqref{fint1}. In tunnel junctions with small transparencies
${\cal D}\ll 1$ one gets $g_{12},\eta_1,\eta_2
\propto{\cal D}$,\; $h_{12},f_{12}\propto {\cal D}^2$.

For the order parameter $f(x)e^{i\chi(x)}$ normalized to $f=1$ in the
bulk without superflow, the first integral of the GL equation in the
presence of the supercurrent \cite{Langer1967} takes the form
\begin{equation}
\left(\frac{df}{d\tilde{x}}\right)^2+f^2\!-
\frac{1}{2}f^4\!+\,\dfrac{4\tilde{j}^2}{27f^2}=
2f_{\infty}^2-\dfrac{3}{2}f_{\infty}^4.
\label{gl1d8}
\end{equation}
Here $\tilde{x}=x/\xi$, $\xi=\sqrt{K/|a|}$ is the temperature
dependent superconducting coherence length, $\tilde{j}$ is the
spatially constant normalized current density $\tilde{j}=
{j}/{j_{dp}}=-(3\sqrt{3}/{2})({d \chi}/{d\tilde{x}})f^2$,\;
$j_{dp}=8|e|K^{1/2}|a|^{3/2}/3\sqrt{3}\hbar b$ is the depairing
current in the bulk, and $f_{\infty}$ is the asymptotic value of $f$
in the depth of the superconducting leads.

The boundary conditions (BC) originate from the variation of
${\cal F}_{int}$ and from
the bulk gradient terms integrated by parts. One starts with the BC
in standard linear approximation in
$f_{1}(-0)\equiv f_{10}$ or $f_{2}(+0)\equiv f_{20}$.
Taking real and imaginary parts of the BC for the complex
quantity $f(x)e^{i\chi(x)}$, one finds the following linear BC
and the expression for the supercurrent
\begin{eqnarray}
&\left(d f_i/d\tilde{x}\right)_{0}=(-1)^i\left[
\left(\tilde{g}_{ii}+\tilde{g}_{12}\right)\!f_{i0}-
\tilde{g}_{12}\cos\chi\, f_{\overline{i}0}\right],
\label{bc1}\\
&\tilde{j}=(3\sqrt{3}/2)\tilde{g}_{12}f_{10}f_{20}\sin\chi.
\label{joscur1}
\end{eqnarray}
Here $i=1,2$, $\overline{i}=3-i$. The phase difference of the order
parameters across the interface is $\chi=\chi_{10}-\chi_{20}$, and
$\tilde{g}_{12}=g_{12}\xi(T)/K$, $\tilde{g}_{11}=g_{11}\xi(T)/K$
and $\tilde{g}_{22}=g_{22}\xi(T)/K$ are the effective dimensionless
coefficients.

For tunnel junctions $|\tilde{g}_{12}|\ll 1$. Since the order
parameters near pair-breaking interfaces vary on a scale
$\agt\xi(T)$, one gets from \eqref{bc1} $\tilde{g}_{ii}f_{i0}\alt 1$,
on account of $|\tilde{g}_{12}|f_{i0}\ll 1$. This signifies, in
particular, that for $\tilde{g}_{ii} \gg1$ a strong interfacial pair
breaking $f_{i0}\alt\tilde{g}_{ii}^{-1}\ll 1$ occurs.

\section{Test for a pronounced interfacial pair breaking}

Consider the supercurrent within the second order perturbation
theory in $\tilde{g}_{12}$. Then, according to \eqref{joscur1},
quantities $f_{10}$ and $f_{20}$ should contain the terms of the
zeroth and the first orders of smallness. One takes $x=\pm0$ in
\eqref{gl1d8} and substitutes there \eqref{bc1} and \eqref{joscur1}.
Since the depairing effects in the bulk would contribute to
\eqref{gl1d8} only beginning with the second order terms, within the
given accuracy $f_{\infty}=1$. Then one obtains the
following equations for $f_{10}$ and $f_{20}$
\begin{equation}
f_{i0}^4-2(1+\tilde{g}_{ii}^2)f_{i0}^2+1=
4\tilde{g}_{12}\tilde{g}_{ii}f_{i0}\left(f_{i0}-f_{\overline{i}0}\cos\chi\right).
\label{gleq1st1}
\end{equation}
In the zeroth order in $\tilde{g}_{12}$ the solutions are
\begin{equation}
f^{(0)^2}_{i0}=(1/2)\Bigl(\sqrt{2+\tilde{g}^2_{ii}}-\tilde{g}_{ii}\Bigr)^2,
\quad i=1,2.
\label{f02i}
\end{equation}
Eq. \eqref{f02i} involves two solutions of
\eqref{gleq1st1}. At $g_{ii}>0$ it describes a pair breaking
$f_0<1$, and then $\left|df^{(0)}/d\tilde{x}\right|_0<1/\sqrt{2}$.
At $g_{ii}<0$ an enhanced superconductivity at the boundary $f_0>1$ occurs
\cite{Fink1969,Buzdin1987,Geshkenbein1988,Samokhin1994,Indekeu2007}.
Then the quantity $\left|df^{(0)}/d\tilde{x}\right|_0>\sqrt{2}g_{ii}^2$ can
take large values, and a strong enhancement would induce a characteristic
scale substantially less than $\xi(T)$ of the leads (see Appendix for details).

The first order corrections $f_{0i}\approx f_{0i}^{(0)}+f_{0i}^{(1)}$, which follow from
\eqref{gleq1st1} and \eqref{f02i}, are
\begin{equation}
f_{i0}^{(1)}=-\tilde{g}_{12}
\bigl(f_{i0}^{(0)}-f_{\overline{i}0}^{(0)}\cos\chi\bigr)\Big/\sqrt{2+\tilde{g}_{ii}^2}.
\label{gleq1st2}
\end{equation}
Substituting the order parameters $f_{0i}$ in \eqref{joscur1}, one
finds the first and the second harmonics of
the supercurrent $\tilde{j}=\tilde{j}_{c1}\sin\chi+\tilde{j}_{c2}
\sin2\chi$ in the Josephson tunnel junctions
\begin{multline}
\tilde{j}_{c1}\!=\!
\dfrac{3\sqrt{3}\tilde{g}_{12}}{4}\!
\left(\!\sqrt{2+\tilde{g}_{11}^2}-\tilde{g}_{11}\right)\!
\left(\!\sqrt{2+\tilde{g}_{22}^2}-\tilde{g}_{22}\right)\!\times\\
\times\biggl[1-\dfrac{\tilde{g}_{12}}{\sqrt{2+\tilde{g}_{11}^2}}-
\dfrac{\tilde{g}_{12}}{\sqrt{2+\tilde{g}_{22}^2}}\biggr]_{\,,}
\label{joscur1b1}
\end{multline}
\begin{multline}
\tilde{j}_{c2}=\dfrac{3\sqrt{3}}{8}\tilde{g}^2_{12}
\sum\limits_{i=1}^{2}\dfrac{1}{\sqrt{2+\tilde{g}_{ii}^2}}
\left(\sqrt{2+\tilde{g}_{\overline{i} \overline{i}}^2}-
\tilde{g}_{\overline{i} \overline{i}}\right)^2.
\label{joscur2b1}
\end{multline}

The second harmonic \eqref{joscur2b1} is induced by the
proximity across the interface. At $g_{ii}<0$, the quantity
$\left|g_{ii}\right|$ is here assumed not to be too large to
retain $|\tilde{j}_c|\ll 1$ and $\left|g_{ii}\right|
\ll \sqrt{K\alpha}$. Otherwise, Eqs. \eqref{joscur1b1} and
\eqref{joscur2b1} are applicable at any values of $\tilde{g}_{ii}$
\footnote{The results of Ref. \onlinecite{Barash2012} concern only
the pair breaking effects in symmetric junctions ($f_0<1$), though
the main equation (3) also applies to $f_0>1$\,. In the corresponding
particular case $\tilde{g}_{11}=\tilde{g}_{22}\equiv g_{\delta}>0$
Eqs. \eqref{joscur1b1} and \eqref{joscur2b1} of this paper are
reduced to Eq. (5) of \cite{Barash2012}. It follows from
Eq. \eqref{joscur2b1} that the second harmonic is positive and does not change its
sign under the sign reversal of $g_{ii}$.}. Further, the small second and third terms
in the square brackets in \eqref{joscur1b1} will be neglected.

One finds from \eqref{joscur1b1} and \eqref{joscur2b1} the following
relationship between the amplitudes $\tilde{j}_{c2}$ and
$\tilde{j}_{c1}$:
\begin{equation}
\tilde{j}_{c2}=\dfrac{\tilde{j}^2_{c1}}{6\sqrt{3}}\sum_{i=1,2}
\frac{1}{\sqrt{2+\tilde{g}^2_{ii}}}\left(\sqrt{2+\tilde{g}_{ii}^2}+\tilde{g}_{ii}\right)^2.
\label{joscur12rel1}
\end{equation}

Under the conditions $|\tilde{g}_{ii}|\ll 1,\,\,(i=1,2)$ one can
disregard the interfacial proximity effects. Then in the
original units ${j}_{c1}\propto |\tau|$, ${j}_{c2}
\propto\sqrt{|\tau|}$, while the relative magnitudes of the two
harmonics are described by the equalities
\begin{equation}
j_{c2}j_{\text{dp}}=0.27{j}_{c1}^2,\qquad
j_{c2}=0.7\tilde{g}_{12}{j}_{c1}.
\label{est1}
\end{equation}

Consider now asymmetric junctions with a pronounced interfacial pair
breaking on one side of the interface, when $\left|\tilde{g}_{11}\right|\ll 1$ and
$\tilde{g}_{22}^2\gg 1$, $\tilde{g}_{22}>0$. In $d$-wave junctions this can take place for
interface-to-crystal orientations, which are close to (100) and (110) orientations on the
opposite banks of a smooth interface. Then Eqs. \eqref{joscur1b1} and
\eqref{joscur2b1} are reduced to $\tilde{j}_{c1}\approx
3\sqrt{3}\tilde{g}_{12}\big/(2\sqrt{2}\tilde{g}_{22})$,\,\,
$\tilde{j}_{c2}\approx
3\sqrt{3}\tilde{g}^2_{12}\big/(4\tilde{g}_{22})$\,
while ${j}_{c1}\propto |\tau|^{3/2}$,
${j}_{c2}\propto|\tau|$ in the original units. The relationships between the harmonics
are
\begin{equation}
{j}_{c2}j_{\text{dp}}=0.385\tilde{g}_{22}{j}_{c1}^2,\qquad
{j}_{c2}=0.7\tilde{g}_{12}{j}_{c1}.
\label{ratjc2jc121d045}
\end{equation}

In symmetric junctions with $\tilde{g}_{11}=\tilde{g}_{22}>0$ and
$\tilde{g}_{ii}^2\gg1$ one gets from \eqref{joscur1b1} and \eqref{joscur2b1}
$\tilde{j}_{c1}={3\sqrt{3}\tilde{g}_{12}}\big/{4\tilde{g}_{22}^2}$,\,
$\tilde{j}_{c2}={3\sqrt{3}\tilde{g}^2_{12}}\big/{4\tilde{g}_{22}^3}
$\,.\, Hence, ${j}_{c1}\propto \tau^2$,\, ${j}_{c2}\propto\tau^2$, and
\begin{equation}
{j }_{c2}j_{\text{dp}}= 0.77\,\tilde{g}_{22}\,{j}^2_{c1},\qquad
{j }_{c2}=\bigl({\tilde{g}_{12}}\big/{\tilde{g}_{22}}\bigr)\,{j}_{c1}.
\label{ratjc2jc1sym}
\end{equation}
In symmetric junctions the quantity $\tilde{j}_{c2}/\tilde{j}_{c1}\propto
\tilde{g}_{22}^{-1}$ diminishes with increasing pair breaking \cite{Barash2012}.

It also follows from \eqref{joscur1b1}-\eqref{joscur12rel1} that
$j_{c2}j_{\text{dp}}= 0.136\,\,{j}^2_{c1}$ for $\left|\tilde{g}_{11}
\right|\ll 1$ and $\tilde{g}_{22}^2\gg 1$, $\tilde{g}_{22}<0$. In
symmetric junctions with $\tilde{g}_{ii}^2\gg 1$, $\tilde{g}_{ii}<0$
one gets $j_{c2}j_{\text{dp}}= 0.19\,{j}^2_{c1}\big/
\left|\tilde{g}_{22}\right|^3$. Under the conditions
$\tilde{g}_{11}<0$, $\tilde{g}_{22}>0$, $\tilde{g}_{ii}^2\gg 1$
($i=1,2$) the relation is $j_{c2}j_{\text{dp}}= 0.385\,\,
\tilde{g}_{22}{j}^2_{c1}$.

Comparing \eqref{est1}, \eqref{ratjc2jc121d045} and \eqref{ratjc2jc1sym},
as well as the results for $\tilde{g}_{ii}<0$, one can conclude that the
quantity $|{j}_{c2}|j_{\text{dp}}/{j}^2_{c1}$ always exceeds unity, when
a pronounced interfacial pair breaking $\tilde{g}_{ii}^2\gg1$,
$\tilde{g}_{ii}>0$ takes place on at least one side of the interface. At
$0.4\tilde{g}_{22}\gg 1$, the strong inequality $|{j}_{c2}|j_{\text{dp}}
\gg {j}^2_{c1}$ emerges as a sure sign of the strong interfacial pair breaking.
By contrast, $j_{c2}j_{\text{dp}}/j_{c1}^2$ is substantially less than
unity for the negligibly weak pair breaking or for the enhanced
superconductivity, on both sides of the interface.

Though the specific temperature dependencies, determined above for both
harmonics at different strengths of the pair breaking, could be identified
near $T_c$, there are no striking differences between them. At the same
time, the power-law temperature dependencies of the harmonics
$j_{ci}=j_{ci,0}|\tau|^{\nu_i}$ actually drop out of
\eqref{est1} -- \eqref{ratjc2jc1sym},
together with the dependencies of the effective coupling constants
$\tilde{g}_{il}=\tilde{g}_{il,0}|\tau|^{-1/2}=(g_{il}\xi_0/K)
|\tau|^{-1/2}$ ($i,l=1,2$) and of the depairing current $j_{\text{dp}}
=j_{\text{dp},0}|\tau|^{3/2}$. Hence, \eqref{est1} -- \eqref{ratjc2jc1sym}
are applicable to the ``low-temperature'' amplitudes of the GL
theory, and then the relation $|{j}_{c2,0}|j_{\text{dp},0}\gg
{j}^2_{c1,0}$ will be valid, if at least one of
$\tilde{g}_{ii,0}$ satisfies the condition $\tilde{g}_{ii,0}\gg 1$.
The latter condition is more restrictive than $\tilde{g}_{ii}\gg 1$.
Since the quantities $\tilde{g}_{ii}$ incorporate contributions from a
relatively wide angular interval of quasiparticle momentum directions,
they can be quite large in anisotropically paired superconductors
near $T_c$, but as a rule, decrease substantially when the temperature
goes down \cite{Barash1995}. However, this is generally not the case for
pair breaking effects induced by magnetic boundaries
\cite{Sauls1988,Nazarov2009}.
If $\tilde{g}_{ii,0}\ll 1$ while $\tilde{g}_{ii}\gg 1$, a crossover
from $|{j}_{c2}|j_{\text{dp}}\gg {j}^2_{c1}$ close to $T_c$ to $|{j}_{c2}
|j_{\text{dp}}\alt{j}^2_{c1}$ will show up with decreasing
temperature, as it is described by \eqref{joscur12rel1}.

Assume now $\tilde{g}_{ii,0}\gg 1$. The GL ``low-temperature''
values of the quantities usually exceed their actual values at $T=0$
by about $(2\div3)$ times. This concerns, in particular, the
depairing current:\, $j_{\text{dp},0}/j_{\text{dp}}(T=0)
\approx 2.6$ \cite{Bardeen1962,Kupriyanov1980,Klapwijk1982,%
Golubov2001}. For the standard Josephson current
\cite{Ambegaokar1963,*Ambegaokar1963err}, one obtains $j_{c1}=
j_{c1,0}|\tau|=2\pi^3T_c|\tau|/7\zeta(3)|e|R_N$ near $T_c$ and
$j_{c1}(T=0)=\pi\Delta_0/2|e|R_N$. Hence $j_{c1,0}/j_{c1}(T=0)=
2.66$. Despite the value it would have for analyzing the
experimental results \cite{Dynes2009}, there still is no microscopic
theory for the effects of strong interfacial pair breaking
in a wide temperature range.
If, qualitatively, no dramatic changes of behavior take place and
$\tilde{g}_{ii,0}\gg 1$, the relation $|{j}_{c2}|j_{\text{dp}}\gg
{j}^2_{c1}$ could remain valid with decreasing temperature
below the GL domain of applicability unless anomalous temperature
dependencies, if present, come into play, e.g., due to Andreev
bound states with low energies $\varepsilon_B \ll\Delta_0$.
The temperature dependence of ($|{j}_{c2}|j_{\text{dp}}\big/
{j}^2_{c1}$) in the whole temperature range
is of interest for further theoretical and experimental studies.

\section{Microscopic formula for $\tilde{g}_{12}$}

Microscopic expressions for $\tilde{g}_{12}$ and for
$\tilde{g}_{ii}$ ($i=1,2$) can be obtained by comparing the
Josephson currents of the GL theory with the corresponding
microscopic results near $T_c$. Consider here standard symmetric SIS
tunnel junctions with the negligibly small pair breaking
$|\tilde{g}_{ii}|\ll 1$. Then the GL expression for the first
harmonic should coincide with the microscopic Ambegaokar-Baratoff
formula \cite{Ambegaokar1963,*Ambegaokar1963err} near $T_c$:
$j_{c1}={4|e||a|}g_{12}\big/({\hbar b})
={\pi|\Delta|^2}\big/({4|e|T_cR_N})$\,.
Here $R_N$ is the junction resistance in the normal state. Since
$K=\hbar^2/4m$, $|a|=\alpha|\tau|$ and, in the absence of the pair
breaking, the BCS gap function near $T_c$ is $|\Delta|^2=8\pi^2T_c
(T_c-T)/(7\zeta(3))$, one obtains
\begin{equation}
\tilde{g}_{12}={2\pi^3T_c mb\xi(T)}\big/({7\zeta(3)e^2\hbar\alpha R_{N}}).
\label{g12RN1}
\end{equation}
Eq. \eqref{g12RN1} can be transformed further with the Gor'kov's
microscopic formulas for $b/\alpha$
\cite{Gor'kov1959ar,*Gor'kov1959br} and with the junction
resistance expressed via the averaged transparency
$R_N^{-1}=e^2k_f\overline{\cal D}/4\pi^2\hbar$. Thus for dirty junctions one
obtains $\tilde{g}_{12}=0.75\overline{\cal D}\xi(T)/\ell$, while
for pure junctions $\tilde{g}_{12}=3\pi^2\overline{\cal D}\xi(T)/(14
\zeta(3)\xi_{0})=1.76\overline{\cal D}\xi(T)/\xi_{0}$. Here $\ell$
is the mean free path and $\xi_0=\hbar v_f/\pi T_c$ is the
zero-temperature coherence length. The quantitative microscopic
formulas obtained here for $\tilde{g}_{12}$ agree with the earlier
estimates \cite{Barash2012}. In particular, in dirty superconductors the
ratio $\xi(T)/l$ can easily reach $100$ even at low temperatures.
Hence, for small and moderate transparencies the quantity $\tilde{g}_{12}=
0.75\overline{\cal D}\xi(T)/l$ can vary from vanishingly small values in
the tunneling limit considered in this paper to those well exceeding $100$
near $T_c$, when a substantial anharmonic behavior of the Josephson current
takes place \cite{Barash2012}.

\section{Next order terms in the current}

The initial expression \eqref{joscur1} for the supercurrent can be
generalized to include
the next order terms, which originate from
the phase dependent biquadratic contributions to
\eqref{fint1}. The resulting formula is obtained after replacing
$\tilde{g}_{12}\rightarrow \tilde{g}_{12}+\tilde{\eta}_{1}(|a|/b)
f_{10}^2+\tilde{\eta}_{2}(|a|/b)f_{20}^2$ in \eqref{joscur1} and
adding $\tilde{j}_{f_{12}}=\tilde{j}_{c2,f}\sin2\chi$, where
\begin{equation}
\tilde{j}_{c2,f}=
({3\sqrt{3}}\big/{2})({|a|}\big/{b})\tilde{f}_{12}f_{10}^2f_{20}^2.
\label{joscur2}
\end{equation}
Here $\tilde{f}_{12}=f_{12}\xi(T)/K$, $\tilde{\eta}_{i}=\eta_{i}
\xi(T)/K$, $i=1,2$. Substituting the zeroth order quantities
\eqref{f02i} in \eqref{joscur2}, one obtains
\begin{multline}
\tilde{j}_{c2,f}=({3\sqrt{3}}\big/{8})({|a|}\big/{b})\tilde{f}_{12}
\bigl(\!\sqrt{2+\tilde{g}^2_{11}}-\tilde{g}_{11}\bigr)^2\times\\ \times
\bigl(\!\sqrt{2+\tilde{g}^2_{22}}-\tilde{g}_{22}\bigr)^2 .
\label{joscur21}
\end{multline}

Both contributions to the second harmonic \eqref{joscur2b1} and
\eqref{joscur21} are of the second order in transparency $\propto
{\cal D}^2$, but \eqref{joscur21} also contains an additional small
parameter $|\tau|=(T_c-T)/T_c$, since $|a|=\alpha|\tau|$. This
allows to disregard \eqref{joscur21} in studying the regular problem
assumed above. However, in a number of specific cases the coupling
constant $g_{12}$ can vanish for symmetry reason
\cite{Geshkenbein1986,Yip1990,Sigrist1991}. This concerns, in
particular, the asymmetric junction between identical
$d_{x^2-y^2}$-wave superconductors with exact (100) and (110)
interface-to-crystal orientations on opposite banks of a smooth
plane interlayer \cite{SigristRice1992,Tanaka1994,Yip1995,%
Walker1996,Ostlund1998,Il'ichev1999,Mannhart2002,Golubov2004,%
Lindstrom2003,Lindstrom2006,Golubov2007}. An additional element of
the point symmetry inherent in such a specific system is the reflection
in the $xz$-plane perpendicular to the interface. Free energy should
be invariant under the latter transformation, while the $d_{xy}$-wave
order parameter on one side of the interface changes its sign and
the $d_{x^2-y^2}$-wave order parameter on another side keeps its
value unchanged. Then the expression containing $\left|\Psi_1-
\Psi_2\right|^2$ in \eqref{fint1} is no longer invariant and,
therefore, the coefficients $\tilde{g}_{12}$, $\tilde{\eta}_{1}$ and
$\tilde{\eta}_{2}$ should vanish in the case in question.
By contrast, the term containing $\left|\Psi_1^2-\Psi_2^2\right|^2$
in \eqref{fint1} remains unchanged under the sign reversal of one of the
order parameters and, hence, the coefficient $f_{12}$ can maintain its
regular value.

In reality, the first harmonic $\tilde{j}_{c1}$ remains finite and,
along with $\tilde{j}_{c2,f}$, still represents a substantial part
of the supercurrent, mainly due to interfacial imperfections such as
faceting, roughness, etc \cite{Il'ichev1999,Mannhart2002,
Golubov2004,Lindstrom2003,Lindstrom2006}. Since $|{j}_{c2,f}|\ll
j_{\text{dp}}$, the relation $|{j}_{c1}|\alt|{j}_{c2,f}|$ always
results in the condition $|{j}_{c2,f}|j_{\text{dp}}\gg{j}_{c1}^2$,
which consequently loses its importance in the special case of
strongly suppressed ${g}_{12}$.

\section{Next order terms in the BC}

Let the parameters ${g}_{12}$ and ${g}_{ii}$ ($i=1,2$) be
independent of $T$ near $T_c$. Since $\tilde{g}_{12}, \tilde{g}_{ii}
\propto\xi(T)$, then close to $T_c$ one will get $|\tilde{g}_{12}|\gg 1$
and/or $|\tilde{g}_{ii}|\gg 1$ due to large values of $\xi(T)$.
However, the coupling constants $|g_{12}|$ and $|g_{ii}|$ can
themselves be very small and the temperature range with large
$|\tilde{g}_{12}|$ and/or $|\tilde{g}_{ii}|$ be too narrow. While
the condition $|\tilde{g}_{12}|\ll 1$, resulting in the
tunneling behavior, is assumed throughout this paper, the range of
variations of $\tilde{g}_{ii}$, defined by the strength of
interfacial proximity effects, is quite wide. It contains, for instance,
small values of $|{g}_{ii}|$. For this reason numerical
coefficients of the order of unity, originating from \eqref{gl1d8},
have been kept in \eqref{gleq1st1}-\eqref{joscur12rel1} on an equal
footing with $\tilde{g}_{ii}^2$. However, the additional terms of
the next order of smallness, which come from the BC, can be
comparable with the terms referred to above and should generally
be taken into account.

To clarify the point, let's represent the BC schematically as
$(df_i/d\tilde{x})_{0}\approx\tilde{\cal A}_{i,0}+(|a|/b)
\tilde{\cal A}_{i,1}$ (i=1,2). Here $\tilde{\cal A}_{i,0}=
{\cal A}_{i,0}\xi(T)/K$  is linear in the order parameters and
coincides with the right hand side of \eqref{bc1}.
The correction $\tilde{\cal A}_{i,1}={\cal A}_{i,1}\xi(T)/K$
appears in the BC both from the quartic and biquadratic terms of the
interface free energy \eqref{fint1} and from the weak temperature
dependence of the GL coefficients in ${\cal A}_{i,0}$. Therefore, in
addition, it involves the temperature derivatives of the coefficients.
As \eqref{gl1d8} contains $({df}\big/{d\tilde{x}})^2$, let's consider
$(df_i/d\tilde{x})_{0}^2\approx\tilde{\cal A}_{i,0}^2+2(|a|/b)
\tilde{\cal A}_{i,0}\tilde{\cal A}_{i,1}$. Here it is the crossed product,
which is the next order correction to the $i$-th equation for
the self-consistent order parameters. Since the expression $2(|a|/b)
\tilde{\cal A}_{i0}\tilde{\cal A}_{i1}=2|a|\xi^2(T){\cal A}_{i0}
{\cal A}_{i1}/(bK^2)=2{\cal A}_{i0}{\cal A}_{i1}/(bK)$ depends on
temperature solely via the order parameter amplitudes entering
${\cal A}_{i,1(0)}$, it results in temperature independent
coefficients in the equations for the order parameters.

The corresponding terms should, in general, be taken
into account for the quantitative description of the Josephson current.
However, as the main contribution to \eqref{gl1d8} from
$(df_i/d\tilde{x})_{0}^2$ is quadratic and the correction is
linear in ${\cal A}_{i0}$, for sufficiently large $|{\cal A}_{i0}|$
the correction is negligibly small as compared to ${\cal A}_{i0}^2$.
For sufficiently small $|{\cal A}_{i0}|$ the correction can now also
be disregarded as compared to the coefficients of the order
of unity in \eqref{gl1d8}.

In tunnel junctions, the basic correction of the given origin is
described by the crossed product $4g_{ii}h_{ii}/(Kb)$. In particular,
in the zeroth approximation in the transparency the order parameters are
\begin{equation}
f_{i0}^{(0)^2}=\left[1+\tilde{g}_{ii}^2+
\sqrt{\left(1+\tilde{g}_{ii}^2\right)^2-L_i}\, \right]^{-1},
\label{f02im}
\end{equation}
where $L_i=1-(4g_{ii}h_{ii})/(Kb)$\,and $g_{ii},\,h_{ii}>0$. The quantities $\tilde{g}_{ii}^2
=g_{ii}^2/K|a|$ are implied here and below to involve $g_{ii}^2$ in
the expanded form $g_{ii}^2\approx g_{ii,c}^2+2\tau g_{ii,c}
(dg_{ii}/d\tau)_c$, where weak temperature dependence of $g_{ii}$
near $T_c $ is taken into account in linear in $\tau$ approximation.

The linear in $\tilde{g}_{12}$ first harmonic is obtained by
substituting \eqref{f02im} in \eqref{joscur1}. Calculating also the
second harmonic, one obtains the modified relation between the
second and the first harmonics:
\begin{equation}
\tilde{j}_{c2}=\frac{\tilde{j}_{c1}^2}{3\sqrt{3}}\!\sum\limits_{i=1}^{2}
\dfrac{1}{\sqrt{2+\tilde{g}_{ii}^2}}
\biggl[1+\tilde{g}_{ii}^2+\sqrt{\left(1+\tilde{g}_{ii}^2\right)^2-L_i}\, \biggr].
\label{ratjc2jc121}
\end{equation}

In disregarding the term $(4g_{ii}h_{ii})/(Kb)$, Eqs. \eqref{f02im}
and \eqref{ratjc2jc121} are reduced to the previous ones,
Eqs. \eqref{f02i} and \eqref{joscur12rel1}. Since the two parameters
$\tilde{g}_{ii}=g_{ii}\xi(T)/K$ and $4g_{ii}h_{ii}/(Kb)$ are
independent of each other, the conditions $|\tilde{g}_{ii}|\ll1$
do not generally exclude the special case $4|g_{ii}|h_{ii}/(Kb)\agt 1$.
Then quadratic in $h_{ii}$ corrections can also be noticeable. However,
for sufficiently small $|g_{ii}|$ the opposite conditions $(4|g_{ii}|
h_{ii}/(Kb))^{1/2}\ll 1$ occur and allow to disregard all the
corresponding terms.

For large $\tilde{g}_{ii}$ the term $(4g_{ii}h_{ii})/(Kb)$ becomes
negligibly small in \eqref{ratjc2jc121}, when the temperature
dependent condition $\xi(T)\tilde{g}_{ii}^3\gg4h_{ii}/b$,
in keeping with $\tilde{g}_{ii}\gg 1$, is valid. Also, for one and
the same $g_{ii}$, the right hand side in \eqref{ratjc2jc121} is
always larger than that in \eqref{joscur12rel1}, if $(g_{ii}h_{ii})/
(Kb)>0$. Therefore, the modified formulas do not alter the main
statement of this paper.

In conclusion, a test for identification of a pronounced
interfacial pair breaking in Josephson tunnel junctions has been
proposed and theoretically verified in this paper, based on
Eqs. \eqref{joscur1b1}-\eqref{joscur12rel1} and \eqref{ratjc2jc121}
obtained within the self-consistent theory of the Josephson current.
The main statement is that the condition $j_{c2}j_{\text{dp}}
\gg{j}_{c1}^2$ indicates to a strong interfacial pair breaking at least
on one side of the interface, if the first and the second harmonics satisfy
the conventional relation $j_{c2}\ll|{j}_{c1}|$.

\begin{acknowledgments}
The support of RFBR grant 11-02-00398 is acknowledged.
\end{acknowledgments}

\appendix*

\section{Order parameter profiles near impenetrable boundaries}

The GL theory allows a detailed description of the spatial profiles of
the order parameters near impenetrable boundaries. Here the
boundaries, which either suppress or enhance the superconductivity in
their vicinities, are considered jointly.

In the case in question the supercurrent vanishes and $f_\infty=1$.
Then Eq. \eqref{gl1d8} reduces to
\begin{equation}
\left(\dfrac{df(\tilde{x})}{d\tilde{x}}\right)^2=
\dfrac12\bigl[1-f^2(\tilde{x})\bigr]^2.
\label{gleq11j02}
\end{equation}

The solution of \eqref{gleq11j02}, which is relevant to the order
parameter near a pair breaking surface at $x=0$ satisfies the
condition $f<1$ throughout the halfspace $x>0$ and takes the form
(see, e.g., Ref. \onlinecite{deGennes1966})
\begin{equation}
f_{\text{pb}}(\tilde{x})=
\tanh\left(\dfrac{\tilde{x}+\tilde{x}_0}{\sqrt{2}}\right).
\label{gleq11j05}
\end{equation}
The parameter $\tilde{x}_0>0$ together with the associated order
parameter value on the surface should be determined from the boundary
conditions.

The expression for the order parameter in superconducting half space
with the pair producing surface directly follows from
\eqref{gleq11j05}, since for each $f(\tilde{x})$, which meets
\eqref{gleq11j02}, the functon $1\big/{f(\tilde{x})}$ satisfies
the same equation \eqref{gleq11j02}. This results in the following
solution
\begin{equation}
f_{\text{pp}}(\tilde{x})=
\coth\left(\dfrac{\tilde{x}+\tilde{x}_0}{\sqrt{2}}\right),
\label{gleq11j08}
\end{equation}\\
for which the condition $f_{pp}>1$ holds throughout the half space $x>0$.

The order parameter $f_{0}$, taken on the surface and described by
\eqref{f02i}, can be alternatively determined by
minimizing full free energy \eqref{fb1} and \eqref{fint1} with the solutions \eqref{gleq11j05} or
\eqref{gleq11j08}. Explicit integration in \eqref{fb1} with  \eqref{gleq11j05} or
\eqref{gleq11j08} results in the part of the bulk free energy modified by the boundary. Retaining
only the quadratic in the order parameter term in the surface free energy \eqref{fint1},
one finds the full free energy per unit square of an impenetrable surface
\begin{equation}
{\cal F}=\dfrac{\sqrt{K}|a|^{3/2}}{\sqrt{2}b}
\left[\dfrac43-2f_0+\dfrac23f_0^3+\sqrt{2}\tilde{g}f_0^2\right],
\label{ffulgpl2}
\end{equation}
for both solutions. The extremum of \eqref{ffulgpl2} does result in
\eqref{f02i} irrespective of the sign of $g$. The surface suppresses the superconducting
order parameter at $g>0$, while at $g<0$ the superconductivity is enhanced near the surface.

It follows from \eqref{gleq11j05} and  \eqref{gleq11j08}
\begin{equation}
\dfrac{df_{\text{pb}}(\tilde{x})}{d\tilde{x}}=\dfrac1{\sqrt{2}}
\sech^2\left(\dfrac{\tilde{x}+\tilde{x}_0}{\sqrt{2}}\right),
\label{gleq11j05dx}
\end{equation}
\begin{equation}
\dfrac{df_{\text{pp}}(\tilde{x})}{d\tilde{x}}=-\dfrac1{\sqrt{2}}
\csch^2\left(\dfrac{\tilde{x}+\tilde{x}_0}{\sqrt{2}}\right).
\label{gleq11j08dx}
\end{equation}
As seen from \eqref{gleq11j05dx}, the order parameter \eqref{gleq11j05},
which is suppressed near the boundary, satisfies not only the condition
$f_{pb}(\tilde{x})<1$, but also the relation
\begin{equation}
\left|\dfrac{df_{\text{pb}}(\tilde{x})}{d\tilde{x}}\right|\le\dfrac1{\sqrt{2}}.
\label{dfdxsupp1}
\end{equation}

For the order parameter \eqref{gleq11j08}, which is
enhanced near the boundary, one gets $f_{\text{pp}}(\tilde{x})>1$. According to
\eqref{gleq11j08} and \eqref{gleq11j08dx}, the smaller the parameter
$x_0$, the larger both the order parameter $f_{\text{pp},0}$ and its
derivative $\left|{df_{\text{pp}}}\big/{d\tilde{x}}\right|_0$ taken on
the boundary. A large spatial derivative $\left|{df_{\text{pp}}}\big/
{d\tilde{x}}\right|_0$ corresponds to a small characteristic scale induced
in a superconductor in the vicinity of the surface.

For $g<0$ and $|\tilde{g}|\gg 1$ one gets from \eqref{f02i} and
\eqref{bc1}\, $f_0\approx\sqrt{2}|\tilde{g}|$ and
$\left|{df}\big/d\tilde{x}\right|_0\approx \sqrt{2}\tilde{g}^2$.
Hence, the effective characteristic scale near the surface is
$x_0\sim\xi(T)/|\tilde{g}|$. For the use of the GL theory near
the surface one assumes $x_0\gg \xi_0$. This results in the
condition $|\tilde{g}|\ll 1/\sqrt{\tau}$, i.e.,
$|g|\ll\sqrt{K\alpha}$.

\providecommand{\noopsort}[1]{}\providecommand{\singleletter}[1]{#1}%


\begin{thebibliography}{48}%
\makeatletter
\providecommand \@ifxundefined [1]{%
 \@ifx{#1\undefined}
}%
\providecommand \@ifnum [1]{%
 \ifnum #1\expandafter \@firstoftwo
 \else \expandafter \@secondoftwo
 \fi
}%
\providecommand \@ifx [1]{%
 \ifx #1\expandafter \@firstoftwo
 \else \expandafter \@secondoftwo
 \fi
}%
\providecommand \natexlab [1]{#1}%
\providecommand \enquote  [1]{``#1''}%
\providecommand \bibnamefont  [1]{#1}%
\providecommand \bibfnamefont [1]{#1}%
\providecommand \citenamefont [1]{#1}%
\providecommand \href@noop [0]{\@secondoftwo}%
\providecommand \href [0]{\begingroup \@sanitize@url \@href}%
\providecommand \@href[1]{\@@startlink{#1}\@@href}%
\providecommand \@@href[1]{\endgroup#1\@@endlink}%
\providecommand \@sanitize@url [0]{\catcode `\\12\catcode `\$12\catcode
  `\&12\catcode `\#12\catcode `\^12\catcode `\_12\catcode `\%12\relax}%
\providecommand \@@startlink[1]{}%
\providecommand \@@endlink[0]{}%
\providecommand \url  [0]{\begingroup\@sanitize@url \@url }%
\providecommand \@url [1]{\endgroup\@href {#1}{\urlprefix }}%
\providecommand \urlprefix  [0]{URL }%
\providecommand \Eprint [0]{\href }%
\providecommand \doibase [0]{http://dx.doi.org/}%
\providecommand \selectlanguage [0]{\@gobble}%
\providecommand \bibinfo  [0]{\@secondoftwo}%
\providecommand \bibfield  [0]{\@secondoftwo}%
\providecommand \translation [1]{[#1]}%
\providecommand \BibitemOpen [0]{}%
\providecommand \bibitemStop [0]{}%
\providecommand \bibitemNoStop [0]{.\EOS\space}%
\providecommand \EOS [0]{\spacefactor3000\relax}%
\providecommand \BibitemShut  [1]{\csname bibitem#1\endcsname}%
\let\auto@bib@innerbib\@empty
%</preamble>
\bibitem [{\citenamefont {Buchholtz}\ and\ \citenamefont
  {Zwicknagl}(1981)}]{Buchholtz1981}%
  \BibitemOpen
  \bibfield  {author} {\bibinfo {author} {\bibfnamefont {L.~J.}\ \bibnamefont
  {Buchholtz}}\ and\ \bibinfo {author} {\bibfnamefont {G.}~\bibnamefont
  {Zwicknagl}},\ }\href {\doibase 10.1103/PhysRevB.23.5788} {\bibfield
  {journal} {\bibinfo  {journal} {Phys. Rev. B}\ }\textbf {\bibinfo {volume}
  {23}},\ \bibinfo {pages} {5788} (\bibinfo {year} {1981})}\BibitemShut
  {NoStop}%
\bibitem [{\citenamefont {Barash}\ \emph {et~al.}(1995)\citenamefont {Barash},
  \citenamefont {Galaktionov},\ and\ \citenamefont {Zaikin}}]{Barash1995}%
  \BibitemOpen
  \bibfield  {author} {\bibinfo {author} {\bibfnamefont {Y.~S.}\ \bibnamefont
  {Barash}}, \bibinfo {author} {\bibfnamefont {A.~V.}\ \bibnamefont
  {Galaktionov}}, \ and\ \bibinfo {author} {\bibfnamefont {A.~D.}\ \bibnamefont
  {Zaikin}},\ }\href {\doibase 10.1103/PhysRevB.52.665} {\bibfield  {journal}
  {\bibinfo  {journal} {Phys. Rev. B}\ }\textbf {\bibinfo {volume} {52}},\
  \bibinfo {pages} {665} (\bibinfo {year} {1995})}\BibitemShut {NoStop}%
\bibitem [{\citenamefont {Matsumoto}\ and\ \citenamefont
  {Shiba}(1995{\natexlab{a}})}]{Shiba1995a}%
  \BibitemOpen
  \bibfield  {author} {\bibinfo {author} {\bibfnamefont {M.}~\bibnamefont
  {Matsumoto}}\ and\ \bibinfo {author} {\bibfnamefont {H.}~\bibnamefont
  {Shiba}},\ }\href {\doibase 10.1143/JPSJ.64.1703} {\bibfield  {journal}
  {\bibinfo  {journal} {J. Phys. Soc. Jpn.}\ }\textbf {\bibinfo {volume}
  {64}},\ \bibinfo {pages} {1703} (\bibinfo {year}
  {1995}{\natexlab{a}})}\BibitemShut {NoStop}%
\bibitem [{\citenamefont {Matsumoto}\ and\ \citenamefont
  {Shiba}(1995{\natexlab{b}})}]{Shiba1995b}%
  \BibitemOpen
  \bibfield  {author} {\bibinfo {author} {\bibfnamefont {M.}~\bibnamefont
  {Matsumoto}}\ and\ \bibinfo {author} {\bibfnamefont {H.}~\bibnamefont
  {Shiba}},\ }\href {\doibase 10.1143/JPSJ.64.3384} {\bibfield  {journal}
  {\bibinfo  {journal} {J. Phys. Soc. Jpn.}\ }\textbf {\bibinfo {volume}
  {64}},\ \bibinfo {pages} {3384} (\bibinfo {year}
  {1995}{\natexlab{b}})}\BibitemShut {NoStop}%
\bibitem [{\citenamefont {Matsumoto}\ and\ \citenamefont
  {Shiba}(1995{\natexlab{c}})}]{Shiba1995c}%
  \BibitemOpen
  \bibfield  {author} {\bibinfo {author} {\bibfnamefont {M.}~\bibnamefont
  {Matsumoto}}\ and\ \bibinfo {author} {\bibfnamefont {H.}~\bibnamefont
  {Shiba}},\ }\href {\doibase 10.1143/JPSJ.64.4867} {\bibfield  {journal}
  {\bibinfo  {journal} {J. Phys. Soc. Jpn.}\ }\textbf {\bibinfo {volume}
  {64}},\ \bibinfo {pages} {4867} (\bibinfo {year}
  {1995}{\natexlab{c}})}\BibitemShut {NoStop}%
\bibitem [{\citenamefont {Matsumoto}\ and\ \citenamefont
  {Shiba}(1996)}]{Shiba1996}%
  \BibitemOpen
  \bibfield  {author} {\bibinfo {author} {\bibfnamefont {M.}~\bibnamefont
  {Matsumoto}}\ and\ \bibinfo {author} {\bibfnamefont {H.}~\bibnamefont
  {Shiba}},\ }\href {\doibase 10.1143/JPSJ.65.2194} {\bibfield  {journal}
  {\bibinfo  {journal} {J. Phys. Soc. Jpn.}\ }\textbf {\bibinfo {volume}
  {65}},\ \bibinfo {pages} {2194} (\bibinfo {year} {1996})}\BibitemShut
  {NoStop}%
\bibitem [{\citenamefont {Nagato}\ and\ \citenamefont
  {Nagai}(1995)}]{Nagai1995}%
  \BibitemOpen
  \bibfield  {author} {\bibinfo {author} {\bibfnamefont {Y.}~\bibnamefont
  {Nagato}}\ and\ \bibinfo {author} {\bibfnamefont {K.}~\bibnamefont {Nagai}},\
  }\href {\doibase 10.1103/PhysRevB.51.16254} {\bibfield  {journal} {\bibinfo
  {journal} {Phys. Rev. B}\ }\textbf {\bibinfo {volume} {51}},\ \bibinfo
  {pages} {16254} (\bibinfo {year} {1995})}\BibitemShut {NoStop}%
\bibitem [{\citenamefont {Buchholtz}\ \emph
  {et~al.}(1995{\natexlab{a}})\citenamefont {Buchholtz}, \citenamefont
  {Palumbo}, \citenamefont {Rainer},\ and\ \citenamefont {Sauls}}]{Sauls1995a}%
  \BibitemOpen
  \bibfield  {author} {\bibinfo {author} {\bibfnamefont {L.~J.}\ \bibnamefont
  {Buchholtz}}, \bibinfo {author} {\bibfnamefont {M.}~\bibnamefont {Palumbo}},
  \bibinfo {author} {\bibfnamefont {D.}~\bibnamefont {Rainer}}, \ and\ \bibinfo
  {author} {\bibfnamefont {J.~A.}\ \bibnamefont {Sauls}},\ }\href@noop {}
  {\bibfield  {journal} {\bibinfo  {journal} {J. Low Temp. Phys.}\ }\textbf
  {\bibinfo {volume} {101}},\ \bibinfo {pages} {1079} (\bibinfo {year}
  {1995}{\natexlab{a}})}\BibitemShut {NoStop}%
\bibitem [{\citenamefont {Buchholtz}\ \emph
  {et~al.}(1995{\natexlab{b}})\citenamefont {Buchholtz}, \citenamefont
  {Palumbo}, \citenamefont {Rainer},\ and\ \citenamefont {Sauls}}]{Sauls1995b}%
  \BibitemOpen
  \bibfield  {author} {\bibinfo {author} {\bibfnamefont {L.~J.}\ \bibnamefont
  {Buchholtz}}, \bibinfo {author} {\bibfnamefont {M.}~\bibnamefont {Palumbo}},
  \bibinfo {author} {\bibfnamefont {D.}~\bibnamefont {Rainer}}, \ and\ \bibinfo
  {author} {\bibfnamefont {J.~A.}\ \bibnamefont {Sauls}},\ }\href
  {http://dx.doi.org/10.1007/BF00754526} {\bibfield  {journal} {\bibinfo
  {journal} {J. Low Temp. Phys.}\ }\textbf {\bibinfo {volume} {101}},\ \bibinfo
  {pages} {1099} (\bibinfo {year} {1995}{\natexlab{b}})}\BibitemShut {NoStop}%
\bibitem [{\citenamefont {Alber}\ \emph {et~al.}(1996)\citenamefont {Alber},
  \citenamefont {B\"auml}, \citenamefont {Ernst}, \citenamefont {Kienle},
  \citenamefont {Kopf},\ and\ \citenamefont {Rouchal}}]{Alber1996}%
  \BibitemOpen
  \bibfield  {author} {\bibinfo {author} {\bibfnamefont {M.}~\bibnamefont
  {Alber}}, \bibinfo {author} {\bibfnamefont {B.}~\bibnamefont {B\"auml}},
  \bibinfo {author} {\bibfnamefont {R.}~\bibnamefont {Ernst}}, \bibinfo
  {author} {\bibfnamefont {D.}~\bibnamefont {Kienle}}, \bibinfo {author}
  {\bibfnamefont {A.}~\bibnamefont {Kopf}}, \ and\ \bibinfo {author}
  {\bibfnamefont {M.}~\bibnamefont {Rouchal}},\ }\href {\doibase
  10.1103/PhysRevB.53.5863} {\bibfield  {journal} {\bibinfo  {journal} {Phys.
  Rev. B}\ }\textbf {\bibinfo {volume} {53}},\ \bibinfo {pages} {5863}
  (\bibinfo {year} {1996})}\BibitemShut {NoStop}%
\bibitem [{\citenamefont {Agterberg}(1997)}]{Agterberg1997}%
  \BibitemOpen
  \bibfield  {author} {\bibinfo {author} {\bibfnamefont {D.~F.}\ \bibnamefont
  {Agterberg}},\ }\href@noop {} {\bibfield  {journal} {\bibinfo  {journal} {J.
  Phys. Condens. Matter}\ }\textbf {\bibinfo {volume} {9}},\ \bibinfo {pages}
  {7435} (\bibinfo {year} {1997})}\BibitemShut {NoStop}%
\bibitem [{\citenamefont {Tokuyasu}\ \emph {et~al.}(1988)\citenamefont
  {Tokuyasu}, \citenamefont {Sauls},\ and\ \citenamefont {Rainer}}]{Sauls1988}%
  \BibitemOpen
  \bibfield  {author} {\bibinfo {author} {\bibfnamefont {T.}~\bibnamefont
  {Tokuyasu}}, \bibinfo {author} {\bibfnamefont {J.~A.}\ \bibnamefont {Sauls}},
  \ and\ \bibinfo {author} {\bibfnamefont {D.}~\bibnamefont {Rainer}},\ }\href
  {\doibase 10.1103/PhysRevB.38.8823} {\bibfield  {journal} {\bibinfo
  {journal} {Phys. Rev. B}\ }\textbf {\bibinfo {volume} {38}},\ \bibinfo
  {pages} {8823} (\bibinfo {year} {1988})}\BibitemShut {NoStop}%
\bibitem [{\citenamefont {Cottet}\ \emph {et~al.}(2009)\citenamefont {Cottet},
  \citenamefont {Huertas-Hernando}, \citenamefont {Belzig},\ and\ \citenamefont
  {Nazarov}}]{Nazarov2009}%
  \BibitemOpen
  \bibfield  {author} {\bibinfo {author} {\bibfnamefont {A.}~\bibnamefont
  {Cottet}}, \bibinfo {author} {\bibfnamefont {D.}~\bibnamefont
  {Huertas-Hernando}}, \bibinfo {author} {\bibfnamefont {W.}~\bibnamefont
  {Belzig}}, \ and\ \bibinfo {author} {\bibfnamefont {Y.~V.}\ \bibnamefont
  {Nazarov}},\ }\href {\doibase 10.1103/PhysRevB.80.184511} {\bibfield
  {journal} {\bibinfo  {journal} {Phys. Rev. B}\ }\textbf {\bibinfo {volume}
  {80}},\ \bibinfo {pages} {184511} (\bibinfo {year} {2009})}\BibitemShut
  {NoStop}%
\bibitem [{\citenamefont {Kimura}\ \emph {et~al.}(2009)\citenamefont {Kimura},
  \citenamefont {Barber}, \citenamefont {Ono}, \citenamefont {Ando},\ and\
  \citenamefont {Dynes}}]{Dynes2009}%
  \BibitemOpen
  \bibfield  {author} {\bibinfo {author} {\bibfnamefont {H.}~\bibnamefont
  {Kimura}}, \bibinfo {author} {\bibfnamefont {R.~P.}\ \bibnamefont {Barber}},
  \bibinfo {author} {\bibfnamefont {S.}~\bibnamefont {Ono}}, \bibinfo {author}
  {\bibfnamefont {Y.}~\bibnamefont {Ando}}, \ and\ \bibinfo {author}
  {\bibfnamefont {R.~C.}\ \bibnamefont {Dynes}},\ }\href {\doibase
  10.1103/PhysRevB.80.144506} {\bibfield  {journal} {\bibinfo  {journal} {Phys.
  Rev. B}\ }\textbf {\bibinfo {volume} {80}},\ \bibinfo {pages} {144506}
  (\bibinfo {year} {2009})}\BibitemShut {NoStop}%
\bibitem [{\citenamefont {Il'ichev}\ \emph {et~al.}(1999)\citenamefont
  {Il'ichev}, \citenamefont {Zakosarenko}, \citenamefont {IJsselsteijn},
  \citenamefont {Hoenig}, \citenamefont {Schultze}, \citenamefont {Meyer},
  \citenamefont {Grajcar},\ and\ \citenamefont {Hlubina}}]{Il'ichev1999}%
  \BibitemOpen
  \bibfield  {author} {\bibinfo {author} {\bibfnamefont {E.}~\bibnamefont
  {Il'ichev}}, \bibinfo {author} {\bibfnamefont {V.}~\bibnamefont
  {Zakosarenko}}, \bibinfo {author} {\bibfnamefont {R.~P.~J.}\ \bibnamefont
  {IJsselsteijn}}, \bibinfo {author} {\bibfnamefont {H.~E.}\ \bibnamefont
  {Hoenig}}, \bibinfo {author} {\bibfnamefont {V.}~\bibnamefont {Schultze}},
  \bibinfo {author} {\bibfnamefont {H.-G.}\ \bibnamefont {Meyer}}, \bibinfo
  {author} {\bibfnamefont {M.}~\bibnamefont {Grajcar}}, \ and\ \bibinfo
  {author} {\bibfnamefont {R.}~\bibnamefont {Hlubina}},\ }\href {\doibase
  10.1103/PhysRevB.60.3096} {\bibfield  {journal} {\bibinfo  {journal} {Phys.
  Rev. B}\ }\textbf {\bibinfo {volume} {60}},\ \bibinfo {pages} {3096}
  (\bibinfo {year} {1999})}\BibitemShut {NoStop}%
\bibitem [{\citenamefont {Lindstr\"om}\ \emph {et~al.}(2003)\citenamefont
  {Lindstr\"om}, \citenamefont {Charlebois}, \citenamefont {Tzalenchuk},
  \citenamefont {Ivanov}, \citenamefont {Amin},\ and\ \citenamefont
  {Zagoskin}}]{Lindstrom2003}%
  \BibitemOpen
  \bibfield  {author} {\bibinfo {author} {\bibfnamefont {T.}~\bibnamefont
  {Lindstr\"om}}, \bibinfo {author} {\bibfnamefont {S.~A.}\ \bibnamefont
  {Charlebois}}, \bibinfo {author} {\bibfnamefont {A.~Y.}\ \bibnamefont
  {Tzalenchuk}}, \bibinfo {author} {\bibfnamefont {Z.}~\bibnamefont {Ivanov}},
  \bibinfo {author} {\bibfnamefont {M.~H.~S.}\ \bibnamefont {Amin}}, \ and\
  \bibinfo {author} {\bibfnamefont {A.~M.}\ \bibnamefont {Zagoskin}},\ }\href
  {\doibase 10.1103/PhysRevLett.90.117002} {\bibfield  {journal} {\bibinfo
  {journal} {Phys. Rev. Lett.}\ }\textbf {\bibinfo {volume} {90}},\ \bibinfo
  {pages} {117002} (\bibinfo {year} {2003})}\BibitemShut {NoStop}%
\bibitem [{\citenamefont {Lindstr\"om}\ \emph {et~al.}(2006)\citenamefont
  {Lindstr\"om}, \citenamefont {Johansson}, \citenamefont {Bauch},
  \citenamefont {Stepantsov}, \citenamefont {Lombardi},\ and\ \citenamefont
  {Charlebois}}]{Lindstrom2006}%
  \BibitemOpen
  \bibfield  {author} {\bibinfo {author} {\bibfnamefont {T.}~\bibnamefont
  {Lindstr\"om}}, \bibinfo {author} {\bibfnamefont {J.}~\bibnamefont
  {Johansson}}, \bibinfo {author} {\bibfnamefont {T.}~\bibnamefont {Bauch}},
  \bibinfo {author} {\bibfnamefont {E.}~\bibnamefont {Stepantsov}}, \bibinfo
  {author} {\bibfnamefont {F.}~\bibnamefont {Lombardi}}, \ and\ \bibinfo
  {author} {\bibfnamefont {S.~A.}\ \bibnamefont {Charlebois}},\ }\href
  {\doibase 10.1103/PhysRevB.74.014503} {\bibfield  {journal} {\bibinfo
  {journal} {Phys. Rev. B}\ }\textbf {\bibinfo {volume} {74}},\ \bibinfo
  {pages} {014503} (\bibinfo {year} {2006})}\BibitemShut {NoStop}%
\bibitem [{\citenamefont {Hilgenkamp}\ and\ \citenamefont
  {Mannhart}(2002)}]{Mannhart2002}%
  \BibitemOpen
  \bibfield  {author} {\bibinfo {author} {\bibfnamefont {H.}~\bibnamefont
  {Hilgenkamp}}\ and\ \bibinfo {author} {\bibfnamefont {J.}~\bibnamefont
  {Mannhart}},\ }\href {\doibase 10.1103/RevModPhys.74.485} {\bibfield
  {journal} {\bibinfo  {journal} {Rev. Mod. Phys.}\ }\textbf {\bibinfo {volume}
  {74}},\ \bibinfo {pages} {485} (\bibinfo {year} {2002})}\BibitemShut
  {NoStop}%
\bibitem [{\citenamefont {Golubov}\ \emph {et~al.}(2004)\citenamefont
  {Golubov}, \citenamefont {Kupriyanov},\ and\ \citenamefont
  {Il'ichev}}]{Golubov2004}%
  \BibitemOpen
  \bibfield  {author} {\bibinfo {author} {\bibfnamefont {A.~A.}\ \bibnamefont
  {Golubov}}, \bibinfo {author} {\bibfnamefont {M.~Y.}\ \bibnamefont
  {Kupriyanov}}, \ and\ \bibinfo {author} {\bibfnamefont {E.}~\bibnamefont
  {Il'ichev}},\ }\href {\doibase 10.1103/RevModPhys.76.411} {\bibfield
  {journal} {\bibinfo  {journal} {Rev. Mod. Phys.}\ }\textbf {\bibinfo {volume}
  {76}},\ \bibinfo {pages} {411} (\bibinfo {year} {2004})}\BibitemShut
  {NoStop}%
\bibitem [{\citenamefont {Frolov}\ \emph {et~al.}(2004)\citenamefont {Frolov},
  \citenamefont {Van~Harlingen}, \citenamefont {Oboznov}, \citenamefont
  {Bolginov},\ and\ \citenamefont {Ryazanov}}]{Ryazanov2004}%
  \BibitemOpen
  \bibfield  {author} {\bibinfo {author} {\bibfnamefont {S.~M.}\ \bibnamefont
  {Frolov}}, \bibinfo {author} {\bibfnamefont {D.~J.}\ \bibnamefont
  {Van~Harlingen}}, \bibinfo {author} {\bibfnamefont {V.~A.}\ \bibnamefont
  {Oboznov}}, \bibinfo {author} {\bibfnamefont {V.~V.}\ \bibnamefont
  {Bolginov}}, \ and\ \bibinfo {author} {\bibfnamefont {V.~V.}\ \bibnamefont
  {Ryazanov}},\ }\href {\doibase 10.1103/PhysRevB.70.144505} {\bibfield
  {journal} {\bibinfo  {journal} {Phys. Rev. B}\ }\textbf {\bibinfo {volume}
  {70}},\ \bibinfo {pages} {144505} (\bibinfo {year} {2004})}\BibitemShut
  {NoStop}%
\bibitem [{\citenamefont {Buzdin}(2005)}]{BuzdinPRB2005}%
  \BibitemOpen
  \bibfield  {author} {\bibinfo {author} {\bibfnamefont {A.}~\bibnamefont
  {Buzdin}},\ }\href {\doibase 10.1103/PhysRevB.72.100501} {\bibfield
  {journal} {\bibinfo  {journal} {Phys. Rev. B}\ }\textbf {\bibinfo {volume}
  {72}},\ \bibinfo {pages} {100501} (\bibinfo {year} {2005})}\BibitemShut
  {NoStop}%
\bibitem [{\citenamefont {Goldobin}\ \emph {et~al.}(2007)\citenamefont
  {Goldobin}, \citenamefont {Koelle}, \citenamefont {Kleiner},\ and\
  \citenamefont {Buzdin}}]{Goldobin2007}%
  \BibitemOpen
  \bibfield  {author} {\bibinfo {author} {\bibfnamefont {E.}~\bibnamefont
  {Goldobin}}, \bibinfo {author} {\bibfnamefont {D.}~\bibnamefont {Koelle}},
  \bibinfo {author} {\bibfnamefont {R.}~\bibnamefont {Kleiner}}, \ and\
  \bibinfo {author} {\bibfnamefont {A.}~\bibnamefont {Buzdin}},\ }\href
  {\doibase 10.1103/PhysRevB.76.224523} {\bibfield  {journal} {\bibinfo
  {journal} {Phys. Rev. B}\ }\textbf {\bibinfo {volume} {76}},\ \bibinfo
  {pages} {224523} (\bibinfo {year} {2007})}\BibitemShut {NoStop}%
\bibitem [{\citenamefont {Langer}\ and\ \citenamefont
  {Ambegaokar}(1967)}]{Langer1967}%
  \BibitemOpen
  \bibfield  {author} {\bibinfo {author} {\bibfnamefont {J.~S.}\ \bibnamefont
  {Langer}}\ and\ \bibinfo {author} {\bibfnamefont {V.}~\bibnamefont
  {Ambegaokar}},\ }\href {\doibase 10.1103/PhysRev.164.498} {\bibfield
  {journal} {\bibinfo  {journal} {Phys. Rev.}\ }\textbf {\bibinfo {volume}
  {164}},\ \bibinfo {pages} {498} (\bibinfo {year} {1967})}\BibitemShut
  {NoStop}%
\bibitem [{\citenamefont {Fink}\ and\ \citenamefont {Joiner}(1969)}]{Fink1969}%
  \BibitemOpen
  \bibfield  {author} {\bibinfo {author} {\bibfnamefont {H.~J.}\ \bibnamefont
  {Fink}}\ and\ \bibinfo {author} {\bibfnamefont {W.~C.~H.}\ \bibnamefont
  {Joiner}},\ }\href {\doibase 10.1103/PhysRevLett.23.120} {\bibfield
  {journal} {\bibinfo  {journal} {Phys. Rev. Lett.}\ }\textbf {\bibinfo
  {volume} {23}},\ \bibinfo {pages} {120} (\bibinfo {year} {1969})}\BibitemShut
  {NoStop}%
\bibitem [{\citenamefont {Khlyustikov}\ and\ \citenamefont
  {Buzdin}(1987)}]{Buzdin1987}%
  \BibitemOpen
  \bibfield  {author} {\bibinfo {author} {\bibfnamefont {I.~N.}\ \bibnamefont
  {Khlyustikov}}\ and\ \bibinfo {author} {\bibfnamefont {A.~I.}\ \bibnamefont
  {Buzdin}},\ }\href@noop {} {\bibfield  {journal} {\bibinfo  {journal} {Adv.
  Phys.}\ }\textbf {\bibinfo {volume} {36}},\ \bibinfo {pages} {271} (\bibinfo
  {year} {1987})}\BibitemShut {NoStop}%
\bibitem [{\citenamefont {Geshkenbein}(1988)}]{Geshkenbein1988}%
  \BibitemOpen
  \bibfield  {author} {\bibinfo {author} {\bibfnamefont {V.~B.}\ \bibnamefont
  {Geshkenbein}},\ }\href@noop {} {\bibfield  {journal} {\bibinfo  {journal}
  {Zh. Eksp. Teor. Fiz.}\ }\textbf {\bibinfo {volume} {94}},\ \bibinfo {pages}
  {368} (\bibinfo {year} {1988})},\ \translation{Sov. Phys. JETP \textbf{67},
  2166 (1988)}\BibitemShut {NoStop}%
\bibitem [{\citenamefont {Samokhin}(1994)}]{Samokhin1994}%
  \BibitemOpen
  \bibfield  {author} {\bibinfo {author} {\bibfnamefont {K.~V.}\ \bibnamefont
  {Samokhin}},\ }\href@noop {} {\bibfield  {journal} {\bibinfo  {journal} {Zh.
  Eksp. Teor. Fiz.}\ }\textbf {\bibinfo {volume} {105}},\ \bibinfo {pages}
  {1684} (\bibinfo {year} {1994})},\ \translation{JETP \textbf{78}, 909
  (1994)}\BibitemShut {NoStop}%
\bibitem [{\citenamefont {Kozhevnikov}\ \emph {et~al.}(2007)\citenamefont
  {Kozhevnikov}, \citenamefont {Bael}, \citenamefont {Sahoo}, \citenamefont
  {Temst}, \citenamefont {Haesendonck}, \citenamefont {Vantomme},\ and\
  \citenamefont {Indekeu}}]{Indekeu2007}%
  \BibitemOpen
  \bibfield  {author} {\bibinfo {author} {\bibfnamefont {V.~F.}\ \bibnamefont
  {Kozhevnikov}}, \bibinfo {author} {\bibfnamefont {M.~J.~V.}\ \bibnamefont
  {Bael}}, \bibinfo {author} {\bibfnamefont {P.~K.}\ \bibnamefont {Sahoo}},
  \bibinfo {author} {\bibfnamefont {K.}~\bibnamefont {Temst}}, \bibinfo
  {author} {\bibfnamefont {C.~V.}\ \bibnamefont {Haesendonck}}, \bibinfo
  {author} {\bibfnamefont {A.}~\bibnamefont {Vantomme}}, \ and\ \bibinfo
  {author} {\bibfnamefont {J.~O.}\ \bibnamefont {Indekeu}},\ }\href
  {http://stacks.iop.org/1367-2630/9/i=3/a=075} {\bibfield  {journal} {\bibinfo
   {journal} {New J. Phys.}\ }\textbf {\bibinfo {volume} {9}},\ \bibinfo
  {pages} {75} (\bibinfo {year} {2007})}\BibitemShut {NoStop}%
\bibitem [{Note1()}]{Note1}%
  \BibitemOpen
  \bibinfo {note} {The results of Ref. \protect \rev@citealpnum {Barash2012}
  concern only the pair breaking effects in symmetric junctions ($f_0<1$),
  though the main equation (3) also applies to $f_0>1$\protect \tmspace
  +\thinmuskip {.1667em}. In the corresponding particular case $\protect
  \mathaccentV {tilde}07E{g}_{11}=\protect \mathaccentV
  {tilde}07E{g}_{22}\equiv g_{\delta }>0$ Eqs. \protect \textup {\hbox
  {\mathsurround \z@ \protect \normalfont (\ignorespaces \ref
  {joscur1b1}\unskip \@@italiccorr )}} and \protect \textup {\hbox
  {\mathsurround \z@ \protect \normalfont (\ignorespaces \ref
  {joscur2b1}\unskip \@@italiccorr )}} of this paper are reduced to Eq. (5) of
  Ref. \protect \rev@citealpnum {Barash2012}. It follows from Eq. \protect
  \textup {\hbox {\mathsurround \z@ \protect \normalfont (\ignorespaces \ref
  {joscur2b1}\unskip \@@italiccorr )}} that the second harmonic is positive and
  does not change its sign under the sign reversal of $g_{ii}$.}\BibitemShut
  {Stop}%
\bibitem [{\citenamefont {Barash}(2012)}]{Barash2012}%
  \BibitemOpen
  \bibfield  {author} {\bibinfo {author} {\bibfnamefont {Y.~S.}\ \bibnamefont
  {Barash}},\ }\href {\doibase 10.1103/PhysRevB.85.100503} {\bibfield
  {journal} {\bibinfo  {journal} {Phys. Rev. B}\ }\textbf {\bibinfo {volume}
  {85}},\ \bibinfo {pages} {100503} (\bibinfo {year} {2012})}\BibitemShut
  {NoStop}%
\bibitem [{\citenamefont {Bardeen}(1962)}]{Bardeen1962}%
  \BibitemOpen
  \bibfield  {author} {\bibinfo {author} {\bibfnamefont {J.}~\bibnamefont
  {Bardeen}},\ }\href {\doibase 10.1103/RevModPhys.34.667} {\bibfield
  {journal} {\bibinfo  {journal} {Rev. Mod. Phys.}\ }\textbf {\bibinfo {volume}
  {34}},\ \bibinfo {pages} {667} (\bibinfo {year} {1962})}\BibitemShut
  {NoStop}%
\bibitem [{\citenamefont {Kupriyanov}\ and\ \citenamefont
  {Lukichev}(1980)}]{Kupriyanov1980}%
  \BibitemOpen
  \bibfield  {author} {\bibinfo {author} {\bibfnamefont {M.~Y.}\ \bibnamefont
  {Kupriyanov}}\ and\ \bibinfo {author} {\bibfnamefont {V.~F.}\ \bibnamefont
  {Lukichev}},\ }\href@noop {} {\bibfield  {journal} {\bibinfo  {journal} {Fiz.
  Nizk. Temp.}\ }\textbf {\bibinfo {volume} {6}},\ \bibinfo {pages} {445}
  (\bibinfo {year} {1980})},\ \translation{Sov. J. Low Temp. Phys. \textbf{6},
  210 (1980)}\BibitemShut {NoStop}%
\bibitem [{\citenamefont {Romijn}\ \emph {et~al.}(1982)\citenamefont {Romijn},
  \citenamefont {Klapwijk}, \citenamefont {Renne},\ and\ \citenamefont
  {Mooij}}]{Klapwijk1982}%
  \BibitemOpen
  \bibfield  {author} {\bibinfo {author} {\bibfnamefont {J.}~\bibnamefont
  {Romijn}}, \bibinfo {author} {\bibfnamefont {T.~M.}\ \bibnamefont
  {Klapwijk}}, \bibinfo {author} {\bibfnamefont {M.~J.}\ \bibnamefont {Renne}},
  \ and\ \bibinfo {author} {\bibfnamefont {J.~E.}\ \bibnamefont {Mooij}},\
  }\href {\doibase 10.1103/PhysRevB.26.3648} {\bibfield  {journal} {\bibinfo
  {journal} {Phys. Rev. B}\ }\textbf {\bibinfo {volume} {26}},\ \bibinfo
  {pages} {3648} (\bibinfo {year} {1982})}\BibitemShut {NoStop}%
\bibitem [{\citenamefont {Geers}\ \emph {et~al.}(2001)\citenamefont {Geers},
  \citenamefont {Hesselberth}, \citenamefont {Aarts},\ and\ \citenamefont
  {Golubov}}]{Golubov2001}%
  \BibitemOpen
  \bibfield  {author} {\bibinfo {author} {\bibfnamefont {J.~M.~E.}\
  \bibnamefont {Geers}}, \bibinfo {author} {\bibfnamefont {M.~B.~S.}\
  \bibnamefont {Hesselberth}}, \bibinfo {author} {\bibfnamefont
  {J.}~\bibnamefont {Aarts}}, \ and\ \bibinfo {author} {\bibfnamefont {A.~A.}\
  \bibnamefont {Golubov}},\ }\href {\doibase 10.1103/PhysRevB.64.094506}
  {\bibfield  {journal} {\bibinfo  {journal} {Phys. Rev. B}\ }\textbf {\bibinfo
  {volume} {64}},\ \bibinfo {pages} {094506} (\bibinfo {year}
  {2001})}\BibitemShut {NoStop}%
\bibitem [{\citenamefont {Ambegaokar}\ and\ \citenamefont
  {Baratoff}(1963{\natexlab{a}})}]{Ambegaokar1963}%
  \BibitemOpen
  \bibfield  {author} {\bibinfo {author} {\bibfnamefont {V.}~\bibnamefont
  {Ambegaokar}}\ and\ \bibinfo {author} {\bibfnamefont {A.}~\bibnamefont
  {Baratoff}},\ }\href {\doibase 10.1103/PhysRevLett.10.486} {\bibfield
  {journal} {\bibinfo  {journal} {Phys. Rev. Lett.}\ }\textbf {\bibinfo
  {volume} {10}},\ \bibinfo {pages} {486} (\bibinfo {year}
  {1963}{\natexlab{a}})}\BibitemShut {NoStop}%
\bibitem [{\citenamefont {Ambegaokar}\ and\ \citenamefont
  {Baratoff}(1963{\natexlab{b}})}]{Ambegaokar1963err}%
  \BibitemOpen
  \bibfield  {author} {\bibinfo {author} {\bibfnamefont {V.}~\bibnamefont
  {Ambegaokar}}\ and\ \bibinfo {author} {\bibfnamefont {A.}~\bibnamefont
  {Baratoff}},\ }\href {\doibase 10.1103/PhysRevLett.11.104} {\bibfield
  {journal} {\bibinfo  {journal} {Phys. Rev. Lett.}\ }\textbf {\bibinfo
  {volume} {11}},\ \bibinfo {pages} {104} (\bibinfo {year}
  {1963}{\natexlab{b}})}\BibitemShut {NoStop}%
\bibitem [{\citenamefont {Gor'kov}(1959{\natexlab{a}})}]{Gor'kov1959ar}%
  \BibitemOpen
  \bibfield  {author} {\bibinfo {author} {\bibfnamefont {L.~P.}\ \bibnamefont
  {Gor'kov}},\ }\href@noop {} {\bibfield  {journal} {\bibinfo  {journal} {Zh.
  Eksp. Teor. Fiz.}\ }\textbf {\bibinfo {volume} {36}},\ \bibinfo {pages}
  {1918} (\bibinfo {year} {1959}{\natexlab{a}})},\ \translation{JETP
  \textbf{9}, 1364 (1959)}\BibitemShut {NoStop}%
\bibitem [{\citenamefont {Gor'kov}(1959{\natexlab{b}})}]{Gor'kov1959br}%
  \BibitemOpen
  \bibfield  {author} {\bibinfo {author} {\bibfnamefont {L.~P.}\ \bibnamefont
  {Gor'kov}},\ }\href@noop {} {\bibfield  {journal} {\bibinfo  {journal} {Zh.
  Eksp. Teor. Fiz.}\ }\textbf {\bibinfo {volume} {37}},\ \bibinfo {pages}
  {1407} (\bibinfo {year} {1959}{\natexlab{b}})},\ \translation{JETP
  \textbf{10}, 998 (1960)}\BibitemShut {NoStop}%
\bibitem [{\citenamefont {Geshkenbein}\ and\ \citenamefont
  {Larkin}(1986)}]{Geshkenbein1986}%
  \BibitemOpen
  \bibfield  {author} {\bibinfo {author} {\bibfnamefont {V.~B.}\ \bibnamefont
  {Geshkenbein}}\ and\ \bibinfo {author} {\bibfnamefont {A.~I.}\ \bibnamefont
  {Larkin}},\ }\href@noop {} {\bibfield  {journal} {\bibinfo  {journal} {Pis'ma
  Zh. Eksp. Teor. Fiz.}\ }\textbf {\bibinfo {volume} {43}},\ \bibinfo {pages}
  {306} (\bibinfo {year} {1986})},\ \translation{JETP Lett. \textbf{43}, 395
  (1986)}\BibitemShut {NoStop}%
\bibitem [{\citenamefont {Yip}\ \emph {et~al.}(1990)\citenamefont {Yip},
  \citenamefont {De~Alcantara~Bonfim},\ and\ \citenamefont {Kumar}}]{Yip1990}%
  \BibitemOpen
  \bibfield  {author} {\bibinfo {author} {\bibfnamefont {S.-K.}\ \bibnamefont
  {Yip}}, \bibinfo {author} {\bibfnamefont {O.~F.}\ \bibnamefont
  {De~Alcantara~Bonfim}}, \ and\ \bibinfo {author} {\bibfnamefont
  {P.}~\bibnamefont {Kumar}},\ }\href {\doibase 10.1103/PhysRevB.41.11214}
  {\bibfield  {journal} {\bibinfo  {journal} {Phys. Rev. B}\ }\textbf {\bibinfo
  {volume} {41}},\ \bibinfo {pages} {11214} (\bibinfo {year}
  {1990})}\BibitemShut {NoStop}%
\bibitem [{\citenamefont {Sigrist}\ and\ \citenamefont
  {Ueda}(1991)}]{Sigrist1991}%
  \BibitemOpen
  \bibfield  {author} {\bibinfo {author} {\bibfnamefont {M.}~\bibnamefont
  {Sigrist}}\ and\ \bibinfo {author} {\bibfnamefont {K.}~\bibnamefont {Ueda}},\
  }\href {\doibase 10.1103/RevModPhys.63.239} {\bibfield  {journal} {\bibinfo
  {journal} {Rev. Mod. Phys.}\ }\textbf {\bibinfo {volume} {63}},\ \bibinfo
  {pages} {239} (\bibinfo {year} {1991})}\BibitemShut {NoStop}%
\bibitem [{\citenamefont {Sigrist}\ and\ \citenamefont
  {Rice}(1992)}]{SigristRice1992}%
  \BibitemOpen
  \bibfield  {author} {\bibinfo {author} {\bibfnamefont {M.}~\bibnamefont
  {Sigrist}}\ and\ \bibinfo {author} {\bibfnamefont {T.~M.}\ \bibnamefont
  {Rice}},\ }\href {\doibase 10.1143/JPSJ.61.4283} {\bibfield  {journal}
  {\bibinfo  {journal} {J. Phys. Soc. Jpn.}\ }\textbf {\bibinfo {volume}
  {61}},\ \bibinfo {pages} {4283} (\bibinfo {year} {1992})}\BibitemShut
  {NoStop}%
\bibitem [{\citenamefont {Tanaka}(1994)}]{Tanaka1994}%
  \BibitemOpen
  \bibfield  {author} {\bibinfo {author} {\bibfnamefont {Y.}~\bibnamefont
  {Tanaka}},\ }\href {\doibase 10.1103/PhysRevLett.72.3871} {\bibfield
  {journal} {\bibinfo  {journal} {Phys. Rev. Lett.}\ }\textbf {\bibinfo
  {volume} {72}},\ \bibinfo {pages} {3871} (\bibinfo {year}
  {1994})}\BibitemShut {NoStop}%
\bibitem [{\citenamefont {Yip}(1995)}]{Yip1995}%
  \BibitemOpen
  \bibfield  {author} {\bibinfo {author} {\bibfnamefont {S.}~\bibnamefont
  {Yip}},\ }\href {\doibase 10.1103/PhysRevB.52.3087} {\bibfield  {journal}
  {\bibinfo  {journal} {Phys. Rev. B}\ }\textbf {\bibinfo {volume} {52}},\
  \bibinfo {pages} {3087} (\bibinfo {year} {1995})}\BibitemShut {NoStop}%
\bibitem [{\citenamefont {Walker}\ and\ \citenamefont
  {Luettmer-Strathmann}(1996)}]{Walker1996}%
  \BibitemOpen
  \bibfield  {author} {\bibinfo {author} {\bibfnamefont {M.~B.}\ \bibnamefont
  {Walker}}\ and\ \bibinfo {author} {\bibfnamefont {J.}~\bibnamefont
  {Luettmer-Strathmann}},\ }\href {\doibase 10.1103/PhysRevB.54.588} {\bibfield
   {journal} {\bibinfo  {journal} {Phys. Rev. B}\ }\textbf {\bibinfo {volume}
  {54}},\ \bibinfo {pages} {588} (\bibinfo {year} {1996})}\BibitemShut
  {NoStop}%
\bibitem [{\citenamefont {\"Ostlund}(1998)}]{Ostlund1998}%
  \BibitemOpen
  \bibfield  {author} {\bibinfo {author} {\bibfnamefont {S.}~\bibnamefont
  {\"Ostlund}},\ }\href {\doibase 10.1103/PhysRevB.58.R14757} {\bibfield
  {journal} {\bibinfo  {journal} {Phys. Rev. B}\ }\textbf {\bibinfo {volume}
  {58}},\ \bibinfo {pages} {R14757} (\bibinfo {year} {1998})}\BibitemShut
  {NoStop}%
\bibitem [{\citenamefont {Yokoyama}\ \emph {et~al.}(2007)\citenamefont
  {Yokoyama}, \citenamefont {Sawa}, \citenamefont {Tanaka},\ and\ \citenamefont
  {Golubov}}]{Golubov2007}%
  \BibitemOpen
  \bibfield  {author} {\bibinfo {author} {\bibfnamefont {T.}~\bibnamefont
  {Yokoyama}}, \bibinfo {author} {\bibfnamefont {Y.}~\bibnamefont {Sawa}},
  \bibinfo {author} {\bibfnamefont {Y.}~\bibnamefont {Tanaka}}, \ and\ \bibinfo
  {author} {\bibfnamefont {A.~A.}\ \bibnamefont {Golubov}},\ }\href {\doibase
  10.1103/PhysRevB.75.020502} {\bibfield  {journal} {\bibinfo  {journal} {Phys.
  Rev. B}\ }\textbf {\bibinfo {volume} {75}},\ \bibinfo {pages} {020502}
  (\bibinfo {year} {2007})}\BibitemShut {NoStop}%
\bibitem [{\citenamefont {de~Gennes}(1966)}]{deGennes1966}%
  \BibitemOpen
  \bibfield  {author} {\bibinfo {author} {\bibfnamefont {P.~G.}\ \bibnamefont
  {de~Gennes}},\ }\href@noop {} {\emph {\bibinfo {title} {Superconductivity of
  Metals and Alloys}}}\ (\bibinfo  {publisher} {Addison Wesley Publishing Co,
  Inc.},\ \bibinfo {address} {Reading, MA},\ \bibinfo {year}
  {1966})\BibitemShut {NoStop}%
\end{thebibliography}
\end{document}